\documentclass[12pt,a4,fleqn,onecolumn,oneside]{article} 
\usepackage{amsmath,amsfonts,graphicx,amsbsy}
\usepackage[amssymb]{SIunits}
\usepackage{lineno}
\begin{document}
\title{ \bf Comments on  the recent  velocity measurement of  the muon neutrinos by the OPERA Collaboration}
\author{\bf Jacek Ciborowski$^1$, Jakub Rembieli\'{n}ski$^2$}
\date{\today}
\maketitle

\vspace{1cm}

\begin{center}
$^1$ University of Warsaw, Department of Physics, Ho\.{z}a 69, 00-681 Warszawa, Poland, \\  cib$@$fuw.edu.pl

$^2$ University of Lodz, Department of Theoretical Physics, Pomorska 149/153,\\  90-236 Lodz, Poland, jaremb$@$uni.lodz.pl

\end{center}

\small { {\bf Abstract:} We argue that the  result quoted  by the OPERA Collaboration  cannot be interpreted as  simply related to  the muon neutrino moving at a superluminal velocity from the point of creation at CERN to the point of interaction at LNGS.
 }

\vspace{2cm}

\noindent
The OPERA Collaboration have conducted time-of-flight measurements of the muon  neutrino velocity and found
it exceeding the velocity of light in vacuum, $c$ \cite{cite:OPERA}. The neutrinos were apparently found travelling the distance
of  730~km in time shorter by 60~ns   than   light would  do. While the  detailed examination of their experimental procedure
is ongoing in the community, we  wish to present comments regarding the physical interpretation of the above result.

\noindent
 Hypothetical particles that would move with velocities higher than $c$ have been named tachyons.
The Einstein special relativity theory (SR) is not the proper framework for describing tachyons.
Direct application of the SR formulae to space-like trajectories leads to negative energies as well as causality violation.
As was mentioned many years ago by  Sudarshan~\cite{cite:Sudarshan-1}, a causal description of superluminal
particles demands  absolute simultaneity  for space-time events or,  equivalently,  a preferred frame of reference.
Consequently, such a description must necessarily break the fundamental paradigm of the special relativity,
namely the relativity principle. In common opinion, this must leads to breaking of the Lorentz invariance,
which  is the fundamental symmetry in  physics. However it was shown by one of us~\cite{cite:Kuba-1} that it is possible
to  preserve Lorentz covariance  in a theory with  a preferred frame. This is achieved  by choosing
the absolute synchronisation procedure instead  that of  Einstein's and allows tachyons to  be incorporated in that
framework. The freedom in the use of different synchronisation schemes in SR follows
from the fact that only the velocity  of light over closed paths  can be measured  without any  assumed
synchronisation conventions for clocks. The theory~\cite{cite:Kuba-1} enables  quantisation of
tachyon fields free of causal paradoxes as well as   vacuum instability (i.e. spontaneous particle creation from the vacuum).
This formalism applied to light or slower than light particles is completely equivalent to that of SR with
the Einstein clock synchronisation procedure. It should be stressed then  that a discovery of a tachyon would not
invalidate nor even modify the Einstein's  theory in the subluminal sector, as is notoriously claimed.
The velocity of light in vacuum,  the limiting velocity in Nature, remains such  for both types of particles: massive particles cannot
be accelerated above it and tachyons cannot be decelerated below it.
In the context of  the OPERA measurement it has to be noted  that the theoretical framework for  describing
tachyons~\cite{cite:Kuba-1} can be applied to  a  direct measurement of superluminal velocities
on the classical level in the Einstein synchronisation  scheme  too.  This is relevant  for further discussion since  the GPS system used
in the OPERA experiment works exactly in this scheme of clock synchronisation.

\noindent
Having set the theoretical  background, we now turn to discuss the quantitative result of the OPERA Collaboration.
The measurement for both neutrinos and light involves two  space points in the reference frame of the Earth:
the point of creation of the neutrino  at CERN and its interaction point at LNGS.
Even if the  Earth  was in motion with respect to the hypothetical preferred frame,  this fact  would not influence
the following considerations. As was mentioned  above, the time-of-flight of the neutrino from the point of
production to the point of interaction has been measured using the Einstein synchronisation procedure assuming
the isotropy of propagation of light. The length measurement, $L$ = 730~km, is synchronisation independent.
The  velocity of the neutrino (below we put $c=1$), computed by the time-of-flight method, has also been calculated in the Einstein
synchronisation: $v=1 + 2.48 \times 10^{-5}$ since  in that convention the time interval was determined;  the average energy of the neutrino beam was 17 GeV~\cite{cite:OPERA}.
The energy, $E$, of the tachyonic neutrino is related to its (tachyonic) mass, $\kappa$, and velocity, $v$, through:
\begin{equation}\label{eq:eq1}
E = \frac{\kappa}{\sqrt{v^2 - 1}} \approx \frac{\kappa}{\sqrt{2 \Delta v}},
\end{equation}
where $\Delta v = v-1$.  The above formula is the consequence  of the  Lorentz covariant definition of the four-momentum,
$p_{\mu} = \kappa \,w_{\mu}$, where $w_{\mu}$ is the space-like four-velocity of a tachyon, normalised as: $w_{\mu}w^{\mu} = -1$.
The dispersion relation for tachyons has the following form: $E^2 - \vec{p}^2 = -\kappa^2$.
Formula~(\ref{eq:eq1})  leads to  the following result for the tachyonic mass of the muon neutrino:
$\kappa\approx 120$~MeV. Neither the beam energy ($0$-th component of the  covariant energy-momentum four-vector)
nor the tachyonic mass (a Lorentz invariant) would depend on the synchronisation procedure.
On the other hand, the muon neutrino mass squared  has been determined from the kinematics of the pion decay at rest:
$\pi \rightarrow \mu \nu_{\mu}$. The last quoted measurement is that  of  Assamagan {\em et al.} (1996) which
yielded  $ m^2_{\nu} = -0.016 \pm 0.023$~MeV$^2$~\cite{cite:Assam}. This value muon neutrino mass squared
was determined from the formula:
\begin{equation}
m^2_{\nu} = m^2_{\pi} + m^2_{\mu} - 2 m_{\pi} \sqrt{ m^2_{\mu} +  k^2_{\mu}},
\end{equation}
where $k_{\mu} = 29.79200\pm 0.00011$~MeV  is the muon momentum measured in the spectrometer.
Thus the result of the OPERA Collaboration,  $\kappa^2 \approx 0.014$~GeV$^2$, is entirely incompatible with
 the  measurement of Assamagan {\em et al.}.
We are thus led to a conclusion, with the reservation following below,  that the effect seen by the OPERA
Collaboration cannot be interpreted in terms of the muon neutrino moving faster than light from the point of creation to the point of interaction.
Assuming the tachyonic mass  according to the   result of  Assamagan {\em et al.} one obtains an  illustrative result that
a neutrino with such a mass should arrive $7\times 10^{-14}$~s earlier than light  after a  distance of  730~km,
far beyond any possibility of measurement.
The above  statement is  true assuming that the OPERA Collaboration  detects muon neutrinos that move directly form
the point of creation at CERN to the point of detection at LNGS  without undergoing any processes  in flight.
A  tachyon $t$, however, may decay in number of exotic channels, for example:   $t\rightarrow t \gamma$ and $t\rightarrow t\;t_1\bar{t}_1$,
where $\gamma$ denotes  the photon and $t_1\bar{t}_1$ -- a tachyon-antitachyon pair (including the case
$t\equiv t_1$)~\cite{cite:3t-decay}. It is up  to a detailed analysis whether the latter complex   process might
deliver a clue to  the observed effect with    $t_1$ being  a tachyonic neutrino  with a large tachyonic mass
(i.e. higher velocity at a given energy), subsequently decaying into the muon neutrino that is detected at the end.

\noindent
Last but not least let us mention that we have considered the hypothesis of the tachyonic electron neutrinos
in the context of the mass measurement  using the tritium decay~\cite{cite:Neutrino-tach-1}.
In the quoted  paper we presented calculations of  the shape of the electron energy spectrum that explained
the observed excess of counts  near the end-point. For that reason this hypothesis  will be confronted with the
new measurement  which is in preparation  by the KATRIN Collaboration~\cite{cite:KATRIN}.

\end{document}